\documentclass[%
 reprint,
 superscriptaddress,
 nofootinbib,
nobibnotes,
 twocolumn, 
 amsmath,amssymb,
 aps,
 prb,
]{revtex4-2}
\usepackage{relsize}
\usepackage[colorlinks = true,
            linkcolor = blue,
            urlcolor  = red,
            citecolor = red,
            anchorcolor = blue]{hyperref}
\usepackage[T1]{fontenc}
\usepackage{graphicx}
\usepackage{dcolumn}
\usepackage{color}
\usepackage{amsmath}
\usepackage{array}
\usepackage{MnSymbol}
\usepackage{glossaries}
\newacronym{CDW}{CDW}{charge-density wave}
\newacronym{BOW}{BOW}{bond-order wave}
\newacronym{DQMC}{DQMC}{determinant quantum Monte Carlo}
\newacronym{QMC}{QMC}{Quantum Monte Carlo}
\newacronym{SSH}{SSH}{Su-Schrieffer-Heeger}
\newacronym{HMC}{HMC}{hybrid Monte Carlo}
\newacronym{2D}{2D}{two-dimension}
\newacronym{1D}{1D}{one-dimension}
\newacronym{eph}{$e$-ph}{electron-phonon}
\newacronym{FS}{FS}{Fermi surface}
\newacronym{bSSH}{bSSH}{``bond'' SSH}
\newacronym{oSSH}{oSSH}{``optical'' SSH}
\newacronym{aSSH}{aSSH}{acoustic SSH}

\begin{document}

\preprint{}
\title{A comparative determinant quantum Monte Carlo study of the acoustic and optical variants of the Su–Schrieffer–Heeger model}

\author{Sohan Malkaruge Costa}
\affiliation{Department of Physics and Astronomy, The University of Tennessee, Knoxville, TN 37996, USA}
\affiliation{Institute of Advanced Materials and Manufacturing, The University of Tennessee, Knoxville, TN 37996, USA\looseness=-1} 

\author{Benjamin Cohen-Stead}
\affiliation{Department of Physics and Astronomy, The University of Tennessee, Knoxville, TN 37996, USA}
\affiliation{Institute of Advanced Materials and Manufacturing, The University of Tennessee, Knoxville, TN 37996, USA\looseness=-1} 

\author{Andy~Tanjaroon~Ly}
\affiliation{Department of Physics and Astronomy, The University of Tennessee, Knoxville, TN 37996, USA}
\affiliation{Institute of Advanced Materials and Manufacturing, The University of Tennessee, Knoxville, TN 37996, USA\looseness=-1} 

\author{James Neuhaus}
\affiliation{Department of Physics and Astronomy, The University of Tennessee, Knoxville, TN 37996, USA}
\affiliation{Institute of Advanced Materials and Manufacturing, The University of Tennessee, Knoxville, TN 37996, USA\looseness=-1} 

\author{Steven Johnston}
\affiliation{Department of Physics and Astronomy, The University of Tennessee, Knoxville, TN 37996, USA}
\affiliation{Institute of Advanced Materials and Manufacturing, The University of Tennessee, Knoxville, TN 37996, USA\looseness=-1}

\date{\today}

\begin{abstract}
We compare the acoustic Su–Schrieffer–Heeger (SSH) model with two of its optical variants where the phonons are defined on either on the sites or bonds of the system. First, we discuss how to make fair comparisons between these models in any dimension by ensuring their dimensionless coupling $\lambda$ and relevant phonon energies are the same. We then use determinant quantum Monte Carlo to perform non-perturbative and sign-problem-free simulations of all three models on one-dimensional chains at and away from half-filling. By comparing the results obtained from each model, we demonstrate that the optical and acoustic models produce near identical results within error bars for suitably chosen phonon energies and $\lambda$ at half-filling. In contrast, the bond model has quantitatively different behavior due to its coupling to the ${\bf q} = 0$ phonon mode. These differences also manifest in the total length of the chain, which shrinks for the bond model but not for the acoustic and optical models when $\lambda \ne 0$. Our results have important implications for quantum Monte Carlo modeling of SSH-like interactions, where these models are sometimes regarded as being interchangeable. 
\end{abstract}

\maketitle

\section{Introduction}

Model Hamiltonians like the Holstein~\cite{Holstein1959studies}, Fr{\"o}hlich~\cite{Frohlich1954electrons}, and \gls*{SSH}~\cite{Barisic1970tightbinding, Su1979solitons} models have played a central role in formulating our understanding of strong \gls*{eph} interactions in solids. These models capture essential aspects of the \gls*{eph} problem while also lending themselves more easily to non-perturbative simulations using powerful numerical methods
~\cite{Freericks1993Holstein, Millis1996fermiliquid, Capone1997small, Scalettar1989competition, Jeckelmann1998density, Hohenadler2004quantum, DeFilippis2006validity, Werner2007efficient, Assaad2007diagrammatic,Goodvin2011optical, Grusdt2015renormalization, Weber2015phonon, Greitemann2018lecture, Esterlis2018breakdown, Karakuzu2018solution, Weber2018twodimensional, Boncia2019spectral, Dee2020relative, Bradley2021superconductivity, goetz2023phases}. 
Auxiliary field \gls*{QMC} methods, for example, can simulate these models without a Fermion sign problem, which allows one to obtain numerically exact results down to low temperatures and across a range of model parameters. 

The \gls*{eph} interaction in the Holstein and Fr{\"o}hlich models arises from a coupling between the lattice displacements and the electron density. (This coupling is purely local in the case of the Holstein model and nonlocal in the Fr{\"o}hlich model.) These models couple the lattice's motion to the electron's potential energy. Alternatively, the \gls*{SSH} model couples the lattice displacements to the nearest-neighbor electronic hopping integral $t$, thus modulating the electron's kinetic energy~\cite{Barisic1970tightbinding, Su1979solitons}. This type of \gls*{eph} interaction was first proposed in a model Hamiltonian context by Bari{\v s}i{\'c}, Labb{\'e}, and Friedel in 1970~\cite{Barisic1970tightbinding} to describe phonon-mediated superconductivity in transition metals, and later by Su, Schrieffer, and Heeger to describe polyacetylene~\cite{Su1979solitons}. Nowadays, this microscopic coupling is often referred to as an ``\gls*{SSH}'' or ``Peierls'' coupling due to its importance in the Peierls transition in \gls*{1D}~\cite{Peierls1992more}. 
More recently, \gls*{SSH}-like \gls*{eph} interactions have attracted a significant amount of interest~\cite{Capone1997small, Sengupta2003Peierls, Weber2015excitation, Sous2018light, Zhang2023bipolaronic, Nocera2021bipolaron, banerjee2023ground, Li2022suppressed, Li2020quantum, Xing2021quantum, CohenStead2023hybrid, Feng2022phase, Cai2022robustness, Meier2016observation, Moller2017typeII, Subhajyoti2022Topological, goetz2023phases} due to their potential relevance to high-$T_\mathrm{c}$ superconductivity~\cite{Sous2018light, Li2020quantum, Zhang2023bipolaronic, CohenStead2023hybrid} and topological states of matter~\cite{Meier2016observation, Moller2017typeII, Subhajyoti2022Topological}. 

One can find several variants of the single-band \gls*{SSH} model in the literature (see also Sec.~\ref{sec:methods}). The first is the \gls*{aSSH} model as it was initially proposed~\cite{Barisic1970tightbinding, Su1979solitons}, where an acoustic phonon branch modulates the nearest-neighbor hopping integrals to linear order in the relative distance between the atoms. There are also two common optical variants of the model; the first, which we refer to as the \gls*{oSSH}  model~\cite{Capone1997small}, replaces the acoustic phonons with a dispersionless optical Einstein branch while retaining the acoustic model's \gls*{eph} interaction terms. The second, which we refer to as the \gls*{bSSH} model~\cite{Sengupta2003Peierls}, defines independent harmonic oscillators for each bond in the system, describing the relative displacement between the atoms forming the bond. 
An important distinction among these models is that the displacement of an individual atom simultaneously modulates two neighboring hopping integrals in the acoustic and optical models. In contrast, each hopping integral is modulated independently in the bond model. We will demonstrate that this difference significantly affects how the electrons couple to the ${\bf q} = 0$ modes. 

The \gls*{aSSH} model has traditionally been challenging to simulate because its dispersion vanishes as $\Omega_{\bf q} \sim v_\mathrm{s}|q|$ at the Brillouin zone center, where $v_\mathrm{s}$ is the velocity of sound. \gls*{QMC} methods, for example, face long autocorrelation times when applied to modeling phonon modes with energies $\Omega_{\bf q}/t \ll 1$ \cite{Hohenadler2008autocorrelations}, which has generally prevented simulation of models with low-energy optical and acoustic branches. 
For this reason, many studies have instead focused on the bond and optical \gls*{SSH} models~\cite{Weber2015excitation, Zhang2023bipolaronic, Nocera2021bipolaron, banerjee2023ground, Xing2021quantum, Feng2022phase, Cai2022robustness}. One can also view these models as describing a lattice with a basis where the optical bond-stretching motion of the atoms is naturally expected. The interactions in the \gls*{bSSH} model have the added advantage of being easier to implement in \gls*{QMC} simulations. 

These aspects have led some to focus on the bond \gls*{SSH} model as an effective model for the optical or acoustic versions of the model. 
This viewpoint is supported by a recent study by Weber {\it et al.}~\cite{Weber2015excitation}, who examined the equivalency of the \gls*{aSSH} and \gls*{bSSH} models in a \gls*{1D} chain at half-filling.\footnote{Weber {\it et al.}~\cite{Weber2015excitation} use the term ``optical'' \gls*{SSH} model for what we call the bond \gls*{SSH} model.} In that study, the authors obtained effective $e$-$e$ interactions by integrating the phonons out of the Hamiltonian and then performed continuous time \gls*{QMC} simulations of the resulting effective model. While this approach allowed them to overcome the autocorrelation time problem associated with the acoustic phonons, it sometimes introduces a Fermion sign problem. Nevertheless, Weber {\it et al.}~\cite{Weber2015excitation} were able to compare results for the bond and acoustic models at half-filling and concluded that the models could indeed be mapped onto one another for suitable re-scaling of the \gls*{eph} coupling constant and characteristic phonon energies. However, they did not examine the equivalence of the bond and acoustic models away from half-filling. Nor did they study the optical variant of the \gls*{SSH} model. Therefore, it is an open question whether the three \gls*{SSH} model variants can be regarded as equivalent over more comprehensive ranges of parameter space. 

Here we study the (in)equivalence of the acoustic, bond, and optical \gls*{SSH} models in \gls*{1D} using numerically exact \gls*{DQMC} simulations. To facilitate this study, we employ a \gls*{HMC} sampling scheme~\cite{Beyl2018revisiting, Batrouni2019Langevin, Duane1987hybrid, CohenStead2022fast}, which enables simulations of \gls*{eph} coupled models on large clusters and with physically realistic phonon energies~\cite{CohenStead2020langevin, CohenStead2023hybrid, Bradley2023charge}. Using this approach, we perform numerically exact simulations of all three models down to low temperatures without a Fermion sign problem. At half-filling, we find that the \gls*{aSSH} and \gls*{oSSH} models are equivalent for an adequately defined value of the dimensionless coupling and suitably scaled phonon energies. In contrast, the \gls*{bSSH} model produces qualitatively different results, contrary to the conclusions of  Ref.~\cite{Weber2015excitation}. Away from half-filling, we find that all three models are inequivalent; however, the differences between the \gls*{oSSH} and \gls*{aSSH} models remain small while the larger discrepancies with the \gls*{bSSH} model persist. We also find that the total length of the chains shrinks in our simulations of the \gls*{bSSH} model. This behavior is driven by a kinetic energy lowering mechanism, where every bond contracts by some amount to increase the magnitude of the effective hopping integrals, and is unique to this model. We expect our results to hold in higher dimensions and in the presence of electron correlations and thus have implications for future \gls*{QMC} simulations of models involving SSH-like interactions.  

\section{Methods}\label{sec:methods}
\subsection{Classes of single-band SSH models}
In this section, we provide definitions for the three variants of the SSH model considered in this work: the acoustic, bond, and optical SSH models. For the general discussion, we assume each model is defined on a $D$-dimensional hypercubic lattice with one orbital per unit cell and only nearest-neighbor hopping. 

The Hamiltonian for all three models can be partitioned as 
\begin{equation}\label{eq:Hpartition}
\hat{H} = \hat{H}_e+ \hat{H}_\mathrm{ph} + \hat{H}_{e\textrm{-ph}},
\end{equation}
where $\hat{H}_e$ and $\hat{H}_\mathrm{ph}$ describe the non-interacting electronic and lattice degrees of freedom, respectively, and $\hat{H}_{e\textrm{-ph}}$ describes their coupling. 

In all three cases, the electron degrees of freedom are described using a single-band tight-binding model
\begin{equation}\label{eq:Hel}
    \hat{H}_e =
    -t\sum_{{\bf i}, \nu,\sigma} (\hat{c}^\dagger_{{\bf i}+{\bf a}_\nu,\sigma}\hat{c}^{\phantom\dagger}_{{\bf i},\sigma} + \textrm{h.c.})
    - \mu\sum_{{\bf i},\sigma} \hat{n}_{{\bf i},\sigma}. 
\end{equation}
Here, $\hat{c}^\dagger_{{\bf i},\sigma}$ ($\hat{c}^{\phantom\dagger}_{{\bf i},\sigma}$) creates (annihilates) a spin-$\sigma$ ($=\uparrow,\downarrow$) electron at lattice site ${\bf i}$ and $\hat{n}_{\mathbf{i},\sigma} = \hat{c}^\dagger_{\mathbf{i},\sigma}\hat{c}^{\phantom\dagger}_{\mathbf{i},\sigma}$ is the spin-$\sigma$ electron number operator for site $\mathbf{i}$. The sum over $\nu$ runs over each of the $D$ spatial dimensions, with ${\bf a}_\nu$ a lattice vector with corresponding lattice spacing $a = |{\bf a}_\nu|$. Lastly, $t$ is the nearest neighbor hopping integral, and $\mu$ is the chemical potential. 

In the original \gls*{aSSH} model, the neighboring lattice displacements are coupled by a harmonic potential such that 
\begin{equation}\label{eq:SSH_potential}
\hat{H}_\mathrm{ph}= \sum_{{\bf i},\nu} \left( \frac{\hat{P}_{\mathbf{i},\nu}^2}{2M_{\rm a}}+
\frac{1}{2}K_{\rm a}\left(\hat{X}_{{\bf i}+\mathbf{a}_\nu, \nu}-\hat{X}_{{\bf i},\nu}\right)^2 \right),
\end{equation}
where $\hat{X}_{\mathbf{i},\nu}$ and $\hat{P}_{\mathbf{i},\nu}$ denote the position and momentum operators describing the motion of the atom at site ${\bf i}$ in the direction $\mathbf{a}_\nu$. $K_{\rm a}$ parameterizes the harmonic potential, and $M_{\rm a}$ is the ion mass. The characteristic frequency of the oscillations is then $\Omega_{\rm a} = \sqrt{K_{\rm a}/M_{\rm a}}$. The coupling between neighboring sites in Eq.~\eqref{eq:SSH_potential} results in $D$ acoustic phonon branches labeled by $\nu$, each corresponding to the motion of the ions polarized along one of the $D$ spatial directions. The corresponding dispersion relation is acoustic, with a linear dispersion at the zone center. 

The linear dispersion of the acoustic model can be difficult to simulate using methods like \gls*{QMC}. To overcome this limitation, the \gls*{oSSH} and \gls*{bSSH} models replace the coupled atomic modes with localized Einstein modes
\begin{equation}\label{eq:optical_potential}  
\hat{H}_\mathrm{ph}= \sum_{\mathbf{i},\nu}\bigg( \frac{\hat{P}_{\mathbf{i},\nu}^2}{2M_{\rm (o,b)}}+\frac{1}{2}K_{\rm (o,b)}\hat{X}_{\mathbf{i},\nu}^2 \bigg), 
\end{equation}
where $K_{\rm o} \ (M_{\rm o})$ and $K_{\rm b} \ (M_{\rm b})$ are the spring constant (ion mass) in the \gls*{oSSH} and \gls*{bSSH} models, respectively. The phonon frequency in each case is then given by $\Omega_{\rm (o,b)} = \sqrt{K_{\rm (o,b)}/M_{\rm (o,b)}}$. The difference between these models is that the phonons are understood to live on the sites in the \gls*{oSSH} model and the bonds in the \gls*{bSSH} model. 

The \gls*{oSSH} and \gls*{bSSH} models share the same form for the lattice degrees of freedom $\hat{H}_{\rm ph}$, while the \gls*{aSSH} and \gls*{oSSH} models share the same form for the $e$-ph coupling term $\hat{H}_{e\textrm{-ph}}$ in position space. Specifically, the hopping integral is modulated by a term that is linear in the relative distance between neighboring ions 
\begin{equation}\label{eq:SSH_int}
\hat{H}_{e\textrm{-ph}} = \alpha_{\textrm{(a,o)}}\sum_{\mathbf{i},\nu,\sigma}(\hat{X}_{\mathbf{i}+\mathbf{a}_\nu,\nu}-\hat{X}_{\mathbf{i},\nu}) (\hat{c}^\dagger_{{\bf i}+{\bf a}_\nu,\sigma}\hat{c}^{\phantom\dagger}_{{\bf i},\sigma} + \textrm{h.c.}). 
\end{equation}
Here $\alpha_{\rm a}$ and $\alpha_{\rm o}$ denote the microscopic $e$-ph coupling constant for the \gls*{aSSH} and \gls*{oSSH} models, respectively. In the \gls*{bSSH} model, each phonon mode is instead only associated with a single bond
\begin{equation}
    \hat{H}_{e\textrm{-ph}} = \alpha_{\rm b} \sum_{\mathbf{i},\nu,\sigma} \hat{X}_{\mathbf{i},\nu} (\hat{c}^\dagger_{{\bf i}+{\bf a}_\nu,\sigma}\hat{c}^{\phantom\dagger}_{{\bf i},\sigma} + \textrm{h.c.}), 
\end{equation}
such that each phonon mode modulates only a single hopping integral.

All three of these models can be expressed in momentum space in the generalized form 
($\hbar = 1$) 
\begin{align}\nonumber
    \hat{H} =&\sum_{{\bf k},\sigma} \xi(\mathbf{k}) \hat{c}^\dagger_{{\bf k},\sigma}\hat{c}^{\phantom\dagger}_{{\bf k},\sigma} 
    + \sum_{{\bf q},\nu}\Omega_\nu({\bf q})
    \left(\hat{b}^\dagger_{{\bf q},\nu}\hat{b}^{\phantom\dagger}_{\bf q,\nu}+\tfrac{1}{2}\right)\\
    &+ \frac{1}{\sqrt{N}}\sum_{{\bf k},{\bf q},\nu,\sigma}
    g_\nu({\bf k},{\bf q})\hat{c}^\dagger_{{\bf k}+{\bf q},\sigma}\hat{c}^{\phantom\dagger}_{{\bf k},\sigma}\left(\hat{b}^\dagger_{-{\bf q},\nu}+\hat{b}^{\phantom\dagger}_{{\bf q},\nu}\right), 
\end{align}
where $\xi(\mathbf{k}) = \epsilon(\mathbf{k})-\mu$ and $\epsilon({\bf k}) = -2 t\sum_\nu \cos(k_\nu a)$ is the bare electron dispersion, $\Omega_{\nu}(\mathbf{q})$ is the bare phonon dispersion, and $g_\nu({\bf k},{\bf q})$ is the $e$-ph coupling constant, which depends on both electron momentum ${\bf k}$ and phonon mode momentum ${\bf q}$.

In the case of the \gls*{oSSH} and \gls*{bSSH} models the bare phonon dispersion is simply $\Omega_\nu(\mathbf{q}) = \Omega_{\rm (o,b)}$, whereas in the \gls*{aSSH} model it is
\begin{equation}
    \Omega_\nu(\mathbf{q}) = 2\Omega_a \left\vert \sin\left(\tfrac{q_\nu a}{2}\right)\right\vert.
\end{equation}
Likewise, the specific functional form of $g_\nu({\bf k},{\bf q})$ depends on the model (acoustic vs. optical vs. bond). It is convenient to re-express the momentum-dependent $e$-ph coupling as
\begin{equation}
    g_\nu(\mathbf{k},\mathbf{q}) = g_\nu \cdot f_\nu(\mathbf{k},\mathbf{q}), 
\end{equation}
where $f_\nu(\mathbf{k},\mathbf{q})$ contains the momentum dependence, 
\begin{equation}
    g_\nu = \frac{\alpha}{\sqrt{2 M \Omega_\nu(\mathbf{q}_{\rm ns})}}
\end{equation}
is a constant, and $\mathbf{q}_{\rm ns}$ is the best available nesting wavevector for the non-interacting \gls*{FS}. The momentum-dependent piece for each model is \cite{Zhang2021, Li2011}
\begin{equation}\label{eq:fkq}
    \begin{aligned}
    f_{{\rm a},\nu}(\mathbf{k},\mathbf{q}) &= 4 {\rm i} \ \sqrt{\left|\sin\left(\tfrac{q_\nu a}{2}\right)\right|} \ \cos((k_\nu+q_\nu/2)a)\\
    f_{{\rm o},\nu}(\mathbf{k},\mathbf{q}) &= 4 {\rm i} \ \ \ \ \sin\left(\tfrac{q_\nu a}{2}\right) \ \ \cos((k_\nu+q_\nu/2)a)\\
    f_{{\rm b},\nu}(\mathbf{k},\mathbf{q}) &= 2 \ \ \ \ \ \ \ e^{{\rm i}q_\nu a/2} \ \ \ \ \cos((k_\nu+q_\nu/2)a).
    \end{aligned}
\end{equation}
One can immediately infer that $g_\nu(\mathbf{k},0)=0$ for both the \gls*{aSSH} and \gls*{oSSH} models, whereas this is not the case in the \gls*{bSSH} model.

\subsection{The dimensionless coupling parameter}\label{sec:lambda}
When simulating models with $e$-ph coupling, a key parameter is the dimensionless coupling $\lambda$. This parameter, for example, enters into the superconducting $T_\mathrm{c}$ of a conventional superconductor when treated at the level of BCS or Eliashberg theory~\cite{Allen1983theory, Carbotte1990properties}. 
For a momentum-dependent $e$-ph coupling constant $g(\bf{k},\bf{q})$, $\lambda$ is defined as
\begin{align}\label{eq:lambda}
    \nonumber\lambda &= 2\mathcal{N}(0)\sum_\nu\left\llangle \frac{|g_\nu(\mathbf{k},\mathbf{q})|^2}{\Omega_\nu(\mathbf{q})}\right\rrangle_\mathrm{FS} \\
    \nonumber&= 2\mathcal{N}(0) \sum_\nu \left(\frac{\tfrac{1}{N^2}\sum_{\mathbf{k},\mathbf{q}} \frac{|g_\nu(\mathbf{k},\mathbf{q})|^2}{\Omega_\nu(\mathbf{q})}\delta(\xi(\mathbf{k}+\mathbf{q}))\delta(\xi(\mathbf{k}))}{\tfrac{1}{N^2}\sum_{\mathbf{k},\mathbf{q}}\delta(\xi(\mathbf{k}+\mathbf{q}))\delta(\xi(\mathbf{k}))}\right)\\
    &= \frac{2}{\mathcal{N}(0) N^2}\sum_{\mathbf{k},\mathbf{q},\nu} \frac{|g_\nu(\mathbf{k},\mathbf{q})|^2}{\Omega_\nu(\mathbf{q})}\delta(\xi(\mathbf{k}+\mathbf{q}))\delta(\xi(\mathbf{k})),
\end{align}
where $\mathcal{N}(0) = \tfrac{1}{N}\sum_\mathbf{k} \delta(\xi(\mathbf{k}))$ is the density of states at the \gls*{FS} per spin species and $\llangle \cdot \rrangle_\mathrm{FS}$ denotes a \gls*{FS} average. We will denote $\lambda$ defined in this way as $\lambda^\mathrm{FS}$.

We now consider two simple schemes for approximating $\lambda^\mathrm{FS}$. 
In both approximations, we assume a constant density of states and set $\mathcal{N}(0) \approx W^{-1}$, with $W = 4Dt$ the non-interacting bandwidth. In the first scheme, we additionally remove the $\delta$-functions appearing in Eq.~\eqref{eq:lambda} and perform a simple average over the Brillouin zone such that 
\begin{align}
    \lambda^\mathrm{BZ}&= \frac{2}{W N^2}\sum_{\mathbf{k},\mathbf{q}, \nu} \frac{|g_\nu(\mathbf{k},\mathbf{q})|^2}{\Omega_\nu(\mathbf{q})}.\label{eq:bz_avg}
\end{align}
In the second scheme, we approximate $\lambda^{\rm FS}$ by
\begin{align}
    \nonumber\lambda^{\rm const} &= \frac{2}{W} \sum_\nu \frac{\max\big(|g_\nu(\mathbf{k},\mathbf{q}_{\rm ns})|^2\big)}{\Omega_\nu(\mathbf{q}_{\rm ns})}\\
    &= \frac{2}{W}\sum_\nu\frac{g_\nu^2}{\Omega_\nu(\mathbf{q}_{\rm ns})} \max\big( |f_\nu(\mathbf{k},\mathbf{q}_{\rm ns}) \big)|^2 \big), \label{eq:const_avg}
\end{align}
where $\max(\cdot)$ indicates the maximum of $|f_\nu(\mathbf{k},\mathbf{q}_{\rm ns}) |^2$ as a function of $\mathbf{k}$.

We will show that at half-filling $(\mu = 0)$, the \gls*{aSSH} and \gls*{oSSH} are approximately equivalent when $\lambda^{\rm const}_{\rm a} = \lambda^{\rm const}_{\rm o} = \lambda^{\rm const}_{\rm b}$ and $2\Omega_{\rm a} = \Omega_{\rm o} = \Omega_{\rm b}$ for small dimensionless couplings, while the \gls*{bSSH} model is inequivalent to the other two. The second condition arises from requiring $\Omega(\mathbf{q}_{\rm ns})$ to be the same for all three models.

\subsection{Expressions for a 1D model}\label{sec:1D}
The discussion until this point applies to single-band SSH models defined on $D$-dimensional hypercubic lattices. To test the equivalence of the three models using \gls*{DQMC}, we focus on \gls*{1D} chains with nearest-neighbor hopping. In \gls*{1D}, we fix $|\mathbf{q}_{\rm ns}| = \pi/a$, the nesting wavevector at half-filling, even as we dope the system. Applying this definition to each of the three models results in
$g_{\rm a} = \alpha_{\rm a}/\sqrt{4 M_{\rm a} \Omega_{\rm a}}$, $g_\textrm{o} = \alpha_\textrm{o}/\sqrt{2 M_\textrm{o} \Omega_\textrm{o}}$, and $g_\textrm{b} = \alpha_\textrm{b}/\sqrt{2 M_\textrm{b} \Omega_\textrm{b}}$ for the \gls*{aSSH}, \gls*{oSSH}, and \gls*{bSSH} models, respectively.

Using the appropriate functional forms for $e$-ph coupling, the Brillouin-zone average approximation introduced in 
Eq.~\eqref{eq:bz_avg}
results in 
\begin{equation}
    \begin{aligned}
        \lambda^{\rm BZ}_{\rm a} &= \frac{2\alpha_{\rm a}^2}{M_{\rm a}\Omega_{\rm a}^2 W},\\
        \lambda^{\rm BZ}_{\rm o} &= \frac{4\alpha_{\rm o}^2}{M_{\rm o}\Omega_{\rm o}^2 W},\\
        \lambda^{\rm BZ}_{\rm b} &= \frac{2\alpha_{\rm b}^2}{M_{\rm b}\Omega_{\rm b}^2 W}, 
    \end{aligned}
\end{equation}
for the \gls*{aSSH}, \gls*{oSSH} and \gls*{bSSH} models, respectively. Alternatively, the approximation introduced in 
Eq.~\eqref{eq:const_avg} results in
\begin{equation}
    \begin{aligned}
    \lambda^{\rm const}_{\rm a} &= \frac{4\alpha_{\rm a}^2}{M_{\rm a}\Omega_{\rm a}^2 W},\\
    \lambda^{\rm const}_{\rm o} &= \frac{16\alpha_{\rm o}^2}{M_{\rm o}\Omega_{\rm o}^2 W},\\
    \lambda^{\rm const}_{\rm b} &= \frac{4\alpha_{\rm b}^2}{M_{\rm b}\Omega_{\rm b}^2 W} 
    \end{aligned}
\end{equation}
for each model. We will show that this second approximation results in an approximate equivalence between the \gls*{aSSH} and \gls*{oSSH} models at half-filling.

Finally, when computing $\lambda^\mathrm{FS}$ given by Eq. (\ref{eq:lambda}), we approximate the $\delta$-functions using Lorentzian distributions with a full-width at half-maximum of $\Gamma = 0.01t$ and perform the corresponding momentum sums using $10^3$ $k$-points in the first Brillouin zone.

\subsection{Quantum Monte Carlo}
We studied Eq.~\eqref{eq:Hpartition} using sign-problem free \gls*{DQMC} simulations~\cite{White1989numerical} on \gls*{1D} chains of length $L$. Our implementation uses the \gls*{HMC} method to sample the phonon fields while adopting both Fourier acceleration and time-step splitting to help reduce autocorrelation times \cite{CohenStead2022fast, Beyl2018revisiting, Batrouni2019Langevin}. The \gls*{HMC} updates use forces calculated by evaluating the exact derivative of the total action as it appears in the Monte Carlo weights used in DQMC \cite{Gotz2022valence}. The simulations, therefore, have a computational cost that scales as $O(\beta L^3)$ in \gls*{1D}.
Finally, when simulating the acoustic model, we subtract off the center of mass motion of the lattice $X_\mathrm{cm} \equiv \sum_{i,l}X_i(\tau=l\Delta\tau)$ after every \gls*{HMC} update. However, in practice, we have found that this subtraction doesn't affect any measured quantities other than $\langle X\rangle$ itself.

All simulations were performed with $12$ or $24$ parallel Markov chains, each performing $10^{4}$ warm-up sweeps and $2.5\times 10^{4}$ measurement sweeps, with $1250$ measurements per bin for a total of $20$ measurements of each observable per Markov chain. In all cases, the imaginary time discretization was set to  $\Delta\tau t = 1/20$. 

In DQMC simulations, it is possible to measure the expectation value of a wide variety of correlation functions. To detect bond ordered wave (BOW) correlations, we measure the bond structure factor
\begin{equation}
    S_B(q,\tau) = \frac{1}{L}\sum_{i,r} e^{-{\rm i} qr} \langle \hat{B}_{i+r}(\tau) \hat{B}_{i}(0) \rangle,
\end{equation}
and the corresponding bond susceptibility
\begin{equation}
    \chi_B(q) = \int_0^\beta S(q,\tau) \ d\tau,
\end{equation}
where
\begin{equation}
    \hat{B}_i = \sum_\sigma (\hat{c}_{i+1,\sigma}^\dagger \hat{c}_{i,\sigma}^{\phantom \dagger} + {\rm h.c.})
\end{equation}
is the nearest-neighbor bond operator.

By performing analytic continuation on the electron Green's function using a parameter-free differential evolution algorithm for analytic continuation (DEAC)~\cite{Nichols2022parameter}, we reconstruct the electron spectral function. For the electrons, they are related by 
\begin{align}
    G_{\sigma}(k,\tau) = \int_0^\beta d\omega \ \frac{e^{-\tau\omega} }{1+e^{-\beta\omega}} A(k,\omega).
\end{align}
In the case of phonons, we measure the phonon position correlation function in momentum space
\begin{align}
    C_X(q,\tau) = \frac{1}{L} \sum_{i,r} e^{-{\rm i}qr} \langle \hat{X}_{i+r}(\tau) \hat{X}_i(0) \rangle,
\end{align}
which is related to the standard phonon Green's function $D(q,\tau)$ according to
\begin{equation}
    D(q,\tau) + D(q,\beta-\tau) = 2 M \Omega(q) C_X(q,\tau).
\end{equation}
We then use the phonon Green's function to extract the re-normalized phonon energy
\begin{align}\label{eq:phonon_disp}
    \Omega(q,0) = \sqrt{\Omega^2(q) + \Pi(q,0)},
\end{align}
where $\Pi(q,{\rm i}\nu_n)$ is a function related to the phonon self-energy and $\nu_n = 2\pi n/\beta$ is bosonic Matsubara frequency. This function is related to the phonon Green's function according to
\begin{align}
    D(q, {\rm i}\nu_n) = \frac{2\Omega(q)}{({\rm i} \nu_n)^2-\Omega^2(q)-\Pi(q,{\rm i}\nu_n)}.
\end{align}

\section{Results}\label{sec:results}
\begin{figure}[t]
    \centering
    \includegraphics[width=\columnwidth]{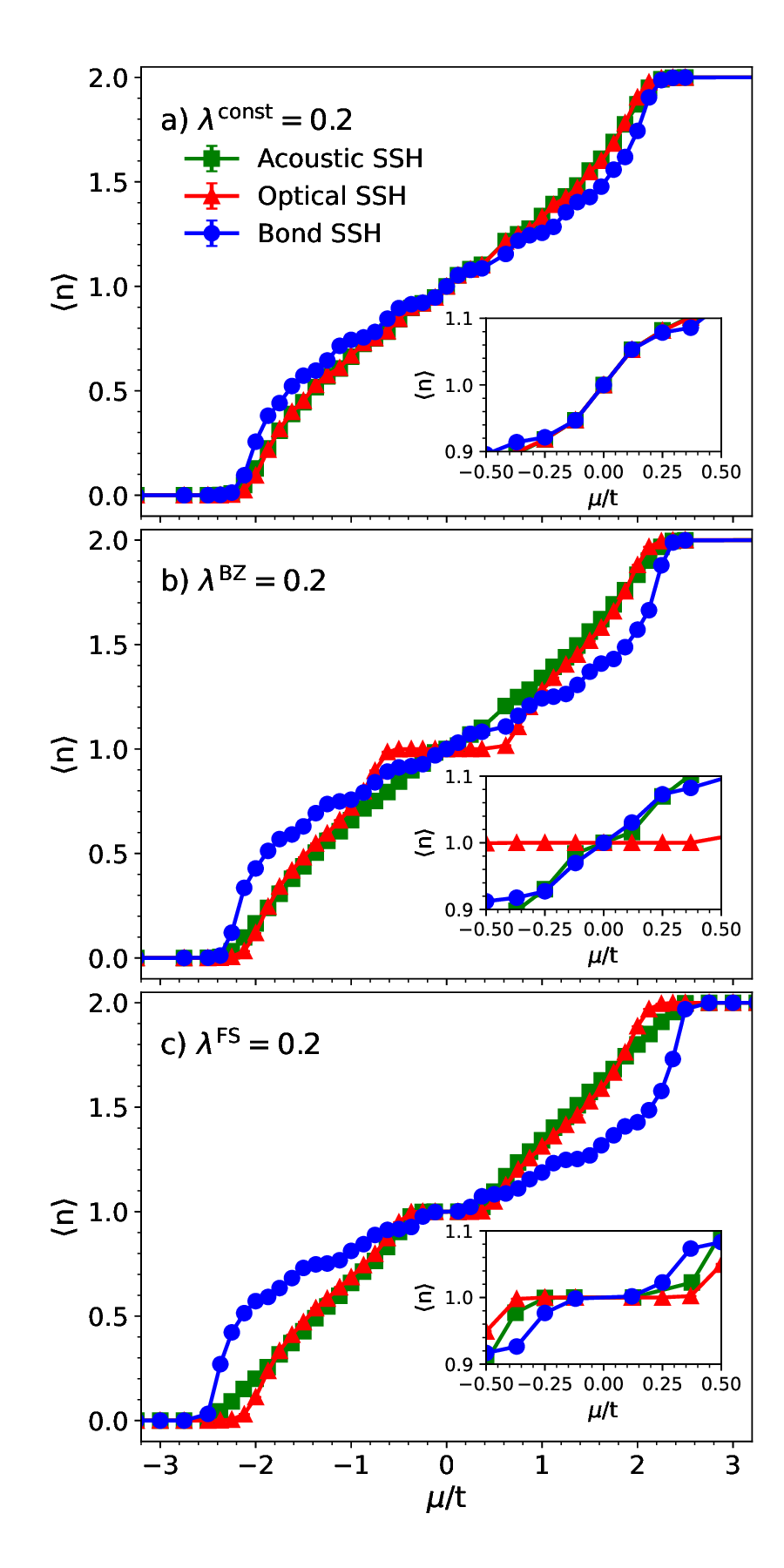}
    \caption{The total filling $\langle n \rangle$ as a function of chemical potential $\mu$ for the acoustic (green $\square$), bond (blue $\bigcirc$), and optical (red $\triangle$) \gls*{SSH} models with fixed a) $\lambda^\mathrm{const} = 0.2$, b) $\lambda^\mathrm{BZ} = 0.2$, and c) $\lambda^\mathrm{FS} = 0.2$. All three panels show results obtained on an $L = 24$ site chain with $\Omega_{\rm b}=\Omega_{\rm o}=2\Omega_{\rm a}=0.1$ and $\beta = 15/t$.}
    \label{fig:nvsmu}
\end{figure}

We will present \gls*{DQMC} results for the models in \gls*{1D} in this section. However, some preliminary comments are in order before proceeding. First, we focus on the weak coupling limit throughout most of this section to avoid complications arising from potential sign changes in the effective hopping integrals. This issue stems from the linear approximation for the \gls*{eph} interaction, which allows the effective distance-dependent hopping to have an unphysical sign change whenever the lattice displacements become sufficiently large~\cite{Nocera2021bipolaron, banerjee2023ground}. Appendix~\ref{sec:sign_change} discusses this issue in greater detail and examines the differences in how the three models approach this limit. 

Second, we will largely focus on results obtained from $L = 24$ site chains throughout this section, even though we have observed some finite-size effects (see Appendix~\ref{sec:finite_site}). (We will present some results for longer $L = 64$ site chains when we examine the spectral functions and renormalized phonon dispersion relations in Secs.~\ref{sec:Akw} and \ref{sec:Bqnu}.) These include an oscillation in the $q = \pi$ bond correlations at half-filling, where the strength of the correlations is under- (over-) predicted relative to the thermodynamic limit for $L = 4n$ ($2n$), where $n$ is an integer. In other words, the bond correlations' strength appears to approach the thermodynamic limit from above or below, depending on whether or not the chain contains an odd number of doubled unit cells associated with the bond-ordered phase. Nevertheless, we have found that the lattice's finite size only affects the quantitative results and not the (in)equivalence of the three \gls*{SSH} models (provided that a consistent chain length is used when making comparisons). Therefore, we will proceed with $L = 24$ site chains, which are rather inexpensive to simulate down to low temperatures. 

\begin{table}[t]
\caption{\label{tbl:coupling} The microscopic values of the \gls*{eph} coupling needed to fix the value of the dimensionless coupling to $0.2$ using the indicated approximations, with $2 \Omega_{\rm a} = \Omega_{\rm o} = \Omega_{\rm b} = 0.1$. The quantity $\lambda_{k_\mathrm{F},2k_\mathrm{F}}\equiv |g(k_\mathrm{F},2k_{\mathrm{F}})|^2/\Omega(2k_\mathrm{F})$ denotes the value of the momentum dependent coupling constant for scattering with $k_\mathrm{F} = \pi/2$ and $2k_\mathrm{F} = \pi$.}
\begin{tabular}{m{0.2\columnwidth}m{0.2\columnwidth}m{0.2\columnwidth}m{0.2\columnwidth}}
\hline\vspace{0.1cm}
Quantity                             & Fixed $\lambda^\mathrm{FS}$ & Fixed $\lambda^\mathrm{BZ}$ & Fixed $\lambda^\mathrm{const}$ \\ \hline\hline
$\alpha_\mathrm{a}$                  & 0.0394 &  0.0316 &  0.0224   \\
$\alpha_\mathrm{o}$                  & 0.0395 &  0.0447 &  0.0224  \\
$\alpha_\mathrm{b}$                  & 0.0787 &  0.0632 &  0.0447 \\
$\lambda_{k_\mathrm{F},2k_\mathrm{F}}^\mathrm{a}$                     & 1.2398 & 0.8000 &  0.4000  \\
$\lambda_{k_\mathrm{F},2k_\mathrm{F}}^\mathrm{o}$                     & 1.2459 & 1.6000 &  0.4000 \\
$\lambda_{k_\mathrm{F},2k_\mathrm{F}}^\mathrm{b}$                     & 1.2398 & 0.8000 &  0.4000 \\ \hline
\end{tabular}
\end{table}

\subsection{Filling vs. chemical potential}
Figure~\ref{fig:nvsmu} plots $\langle n\rangle$ vs. $\mu$ for all three models, where we have fixed the dimensionless coupling $\lambda = 0.2$ using the indicated approximation scheme. For reference, Tbl.~\ref{tbl:coupling} provides the corresponding values of the microscopic \gls*{eph} coupling constants, showing how they differ in each approximation for $\lambda$.

For fixed $\lambda^\mathrm{const} = 0.2$ (see Fig.~\ref{fig:nvsmu}a), $\langle n \rangle$ is a smooth function of $\mu$ for all three models with no clear indications of an energy gap at $\beta = 15/t$. The behavior of the \gls*{aSSH} and \gls*{oSSH} models are identical to within the simulation's error bars for this approximation for the dimensionless coupling. Moreover, the bandwidth inferred from these curves remains equal to the non-interacting value $W = 4t$. On the other hand, the behavior of $\langle n\rangle$ vs. $\mu$ for the \gls*{bSSH} model differs from the acoustic and optical models with an apparent bandwidth increase. This behavior is can also be seen in the single particle spectral functions discussed in Sec.~\ref{sec:Akw}.

Constraining the value of $\lambda^\mathrm{BZ} = 0.2$ (see Fig.~\ref{fig:nvsmu}b) leads to deviations in the $\langle n \rangle$ vs. $\mu$ curves of all three models. The most significant difference between the acoustic and optical models occurs close to half-filling due to the formation of a robust $q = \pi$ bond ordering in the \gls*{oSSH} model (see Sec.~\ref{sec:bond_order}). 
The difference between the \gls*{bSSH} model and both the \gls*{oSSH} and \gls*{aSSH} models also become more pronounced, with the bandwidth of the bond model increasing relative to the results shown in Fig.~\ref{fig:nvsmu}a. Fixing $\lambda^\mathrm{FS} = 0.2$ (see Fig.~\ref{fig:nvsmu}c) results in even more significant deviations between the three models. For this coupling, all three models have a gap associated with the bond order, with the \gls*{oSSH} (\gls*{bSSH}) model having the largest (smallest) gap. Constraining $\lambda^\mathrm{FS} = 0.2$ also further increases the effective bandwidth for the \gls*{bSSH} model, while the other models acquire some widening due to the self-energy broadening of the electronic structure. 

The increasing deviations between the models for the BZ- and FS-averaged values of $\lambda$ stem from differences in how these approximations average the momentum dependence of the coupling constants $g(k,q)$. For weak \gls*{eph} coupling, the physics of the models is dominated by scattering processes across the Fermi surface, i.e., $k_\mathrm{F} = \pi/2$ and $q = 2k_\mathrm{F} = \pi$ at half-filling. Fixing $\lambda^\mathrm{const}$ for the \gls*{1D} model imposes the condition that the mode-resolved dimensionless coupling $\lambda_{k_\mathrm{F},2k_\mathrm{F}} \equiv |g(k_\mathrm{F},2k_\mathrm{F})|^2/\Omega(2k_\mathrm{F})$ be the same for all three models. Conversely, fixing $\lambda^\mathrm{BZ}$ or $\lambda^\mathrm{FS}$, which involves different averages of the momentum-dependent coupling $|g(k,q)|^2$, translates into different values for this quantity, as summarized in Tbl.~\ref{tbl:coupling}. This re-scaling of the effective scattering across the \gls*{FS} also explains the different values of the relative gap sizes inferred in Fig.~\ref{fig:nvsmu}. 

These considerations imply that one should fix $\lambda^\mathrm{const}$ if one wishes to make direct comparisons between the different \gls*{SSH} models in \gls*{1D}. We do not, however, expect this to generally be the case in higher dimensions if the Fermi surface is not perfectly nested; in that case, one should use either BZ- or FS-averaged dimensionless couplings to correctly average over the different scattering processes that enter when the relevant Fermi surfaces are no longer perfectly nested. 

\subsection{Contraction of the lattice in the bond model}\label{sec:deltaL}
\begin{figure}[t]
    \centering
    \includegraphics[width=1.0\columnwidth]{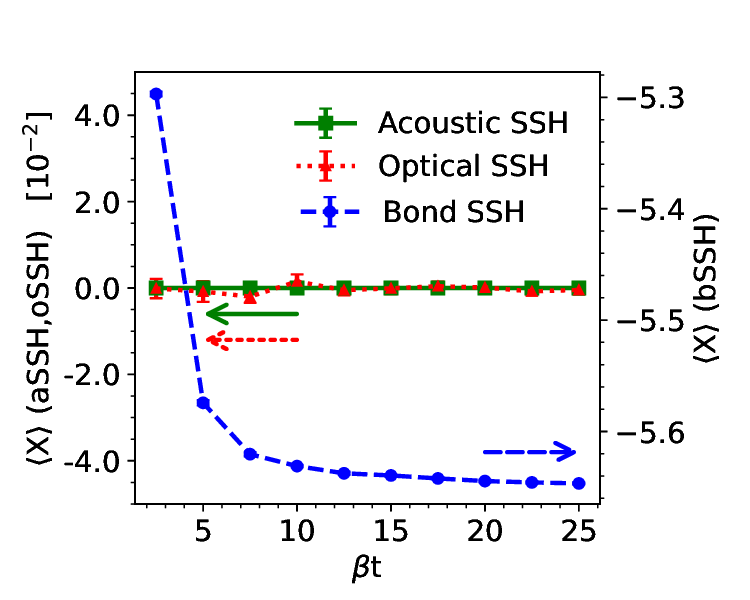}
    \caption{The expectation value of the average phonon displacement 
    $\langle X \rangle$, with a corresponding net change in the length of the lattice $\Delta L = L \langle X \rangle$. Results were obtained here for a half-filled ($\mu = 0$) $L=24$ chain with $\lambda^\mathrm{const}=0.2$ and $\Omega_\mathrm{b}=\Omega_\mathrm{o}=2\Omega_\mathrm{a}=0.1$. The $y$-axis on the left side of the plot is for the acoustic (green $\square$) and optical (red $\triangle$) \gls*{SSH} models. The $y$-axis on the right-hand side of the plot is for the bond \gls*{SSH} model (blue $\bigcirc$). }
    \label{fig:deltaL}
\end{figure}

As noted in the introduction, the \gls*{bSSH} model differs from the acoustic and optical models in how it couples to the $q = 0$ phonon modes. In the acoustic and optical models, the displacement of individual atoms simultaneously shortens one of its neighboring bonds and lengthens the other. This constraint maintains the total length of the chain at all times and decouples the electrons from the ${\bf q} = 0$ phonon mode, which is the reason why $\lim_{q\rightarrow 0}g_{\mathrm{o},\mathrm{a}}(k,q)\rightarrow 0$ [see Eq.~\eqref{eq:fkq}]. We can easily understand this behavior by recognizing that a $q = 0$ mode in the acoustic or optical models translates the chain to the left or right without changing internal bond lengths. The situation is fundamentally different in the bond model. There, the electrons can lower their total kinetic energy by contracting all bonds by the same amount, increasing the effective hopping integrals along each bond. Because of this, $\lim_{q\rightarrow 0} g_{\mathrm{b}}(k,q)\ne 0$ for the bond model [see Eq.~\eqref{eq:fkq}], and the total length of the chain is no longer conserved.  

Figure~\ref{fig:deltaL} confirms these expectations by plotting the net change in the length of the chain, $\Delta L/L$, for all three models as a function of temperature when $\langle n \rangle =1$ and $\lambda^\mathrm{const} = 0.2$. $\Delta L$ is obtained here by averaging the lattice displacements over imaginary time $\Delta L/L \equiv \frac{1}{N_\tau L}\sum_{i,l} X_{i}(l\Delta\tau)$, where $l = 0,\dots, N_\tau$ and $N_\tau$ is the number of imaginary time slices. As expected, $\Delta L$ fluctuates around zero within error bars for the acoustic and optical \gls*{SSH} models at all temperatures. Conversely, it drops from $\Delta L/L = -5.297 \pm 0.002$ at $\beta = 2.5/t$ to $-5.647\pm 0.001$ at $\beta = 25/t$ for the bond model as the thermal fluctuations of the lattice freeze out. Notably, this effect is not limited to \gls*{1D} as we have also observed it in simulations of the bond model in 2D whenever there is a non-zero \gls*{eph} coupling.  

\subsection{Bond order correlations}\label{sec:bond_order}
It is well known that the half-filled \gls*{1D} \gls*{SSH} models are prone to lattice dimerization at low temperatures. This transition, often referred to as the Peierls transition~\cite{Peierls1992more}, is accompanied by $q = \pi$ bond-order correlations. 

\begin{figure}[t]
    \centering
    \includegraphics[width=\columnwidth]{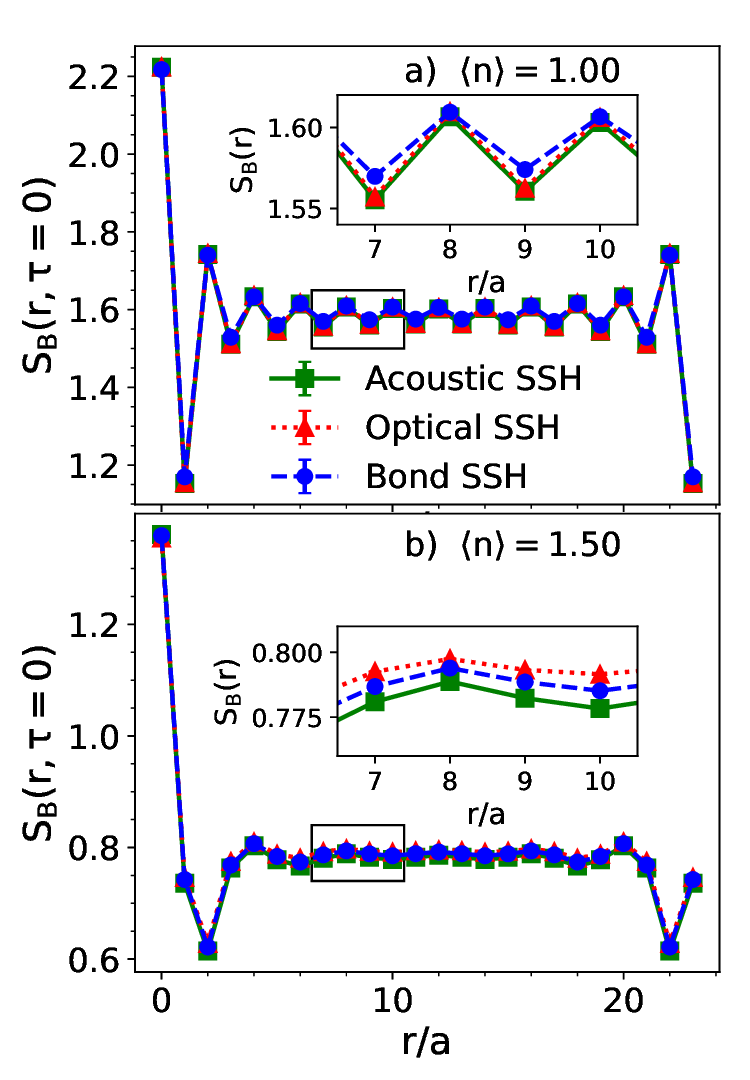}
    \caption{The real space bond-order correlations $\rm S_\mathrm{B}(r)$ as a function of distance along the chain. Results are shown for a) half-filling $\langle n \rangle = 1$ and b) $\langle n \rangle = 1.5$ and were obtained on an $L=24$ site chain with, $\lambda^\mathrm{const}=0.2$, $\Omega_\mathrm{b}=\Omega_\mathrm{o}=2\Omega_\mathrm{a}=0.1$, and $\beta = 15/t$. The insets in both panels provide a zoomed-in view of the region indicated by the small black boxes in the main panels.}
\label{fig:chib_r}
\end{figure}

Figure~\ref{fig:chib_r}a plots the real-space bond correlations obtained for the three  models at half-filling with $\beta = 15/t$, $\lambda^\mathrm{const} = 0.2$, and $\Omega_\mathrm{b}=\Omega_\mathrm{o}=2\Omega_\mathrm{a}=0.1$. For these parameters, all three models develop strong $q = \pi$ correlations that extend along the entire length of the chain. Moreover, the observed correlations for the acoustic and optical models are identical within the error bars. This result further establishes the equivalence of these models after we appropriately fix $\lambda^\mathrm{const}$ and phonon energy scales. The \gls*{bSSH} model also develops comparable bond-order correlations at this temperature, but the overall strength of the modulations is slightly smaller as shown in the inset. Interestingly, the values of the bond model's correlations on the even distances are comparable to those obtained using the other two models. In contrast, the values at odd distances are slightly larger for the bond model, resulting in a weaker amplitude in the overall modulation. 

\begin{figure}[t]
    \centering
    \includegraphics[width=\columnwidth]{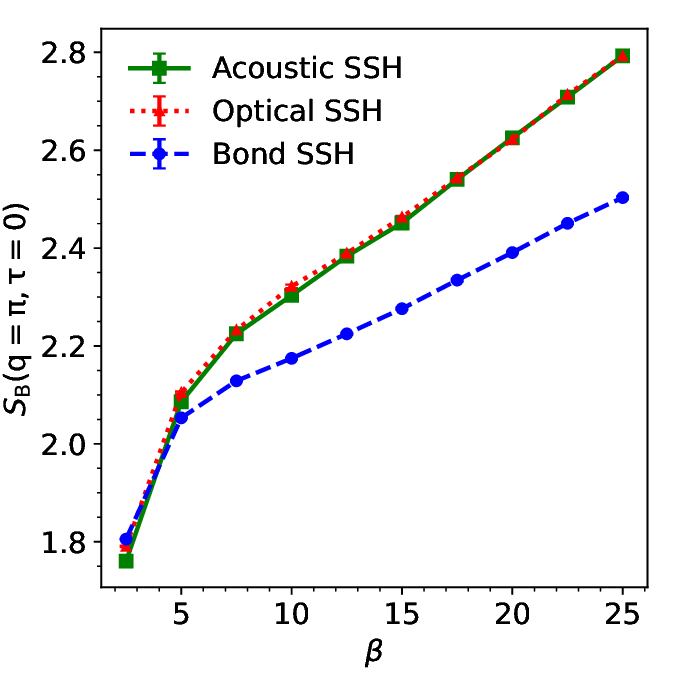}
    \caption{A comparison of the $q = \pi$ bond correlations measured in the acoustic (green $\square$), 
    bond (blue $\bigcirc$), and optical (red $\triangle$) SSH models. Results were obtained    
    $L=24$ site chain at half-filling $\langle n\rangle = 1$ ($\mu = 0$) with $\Omega_\mathrm{b}=\Omega_\mathrm{o}=2\Omega_\mathrm{a}=0.1$ and $\lambda^\mathrm{const} = 0.2$ for all three models.}
    \label{fig:bondcorr_vs_T}
\end{figure}

For comparison, Fig.~\ref{fig:chib_r}b plots the real-space bond correlations for the doped model. In this case, we fix the filling for all three models to $\langle n \rangle = 1.50$ by tuning the chemical potential during the \gls*{DQMC} simulation as described in Ref.~\cite{Miles2022dynamical}. In all three cases, we observe a bond-order correlation with a wave vector $q = 2k_\mathrm{F} = \pi/2$, where $k_\mathrm{F} = 3\pi/4$ 
is the Fermi momentum for the doped system. 
The strength of the modulations is different for each model because the value of $\lambda^\mathrm{const}$ was fixed based on the Fermi surface of the half-filled system. This result demonstrates that any equivalence achieved between the \gls*{aSSH} and \gls*{oSSH} models at half-filling will not persist once the system is doped unless further changes to the model parameters are made. 

\begin{figure}[t]
    \centering
    \includegraphics[width=\columnwidth]{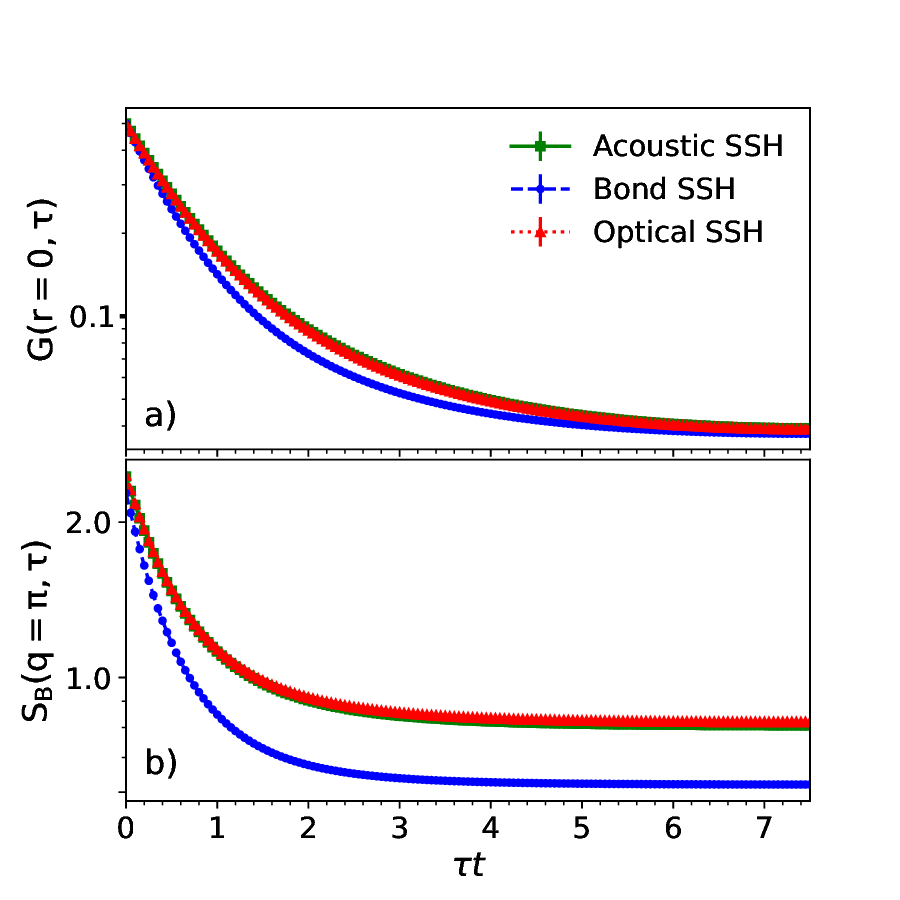}
    \caption{A comparison of the a) local Green's function $G(r=0,\tau)$ and c) bond correlations $S(q = \pi,\tau)$ of the acoustic (green $\square$), bond (blue $\bigcirc$), and optical (red $\triangle$) SSH models. Results were obtained $L=24$ site chain at half-filling $\langle n\rangle = 1$ ($\mu = 0$) with $\Omega_\mathrm{b}=\Omega_\mathrm{o}=2\Omega_\mathrm{a}=0.1$, $\beta = 15/t$, and $\lambda^\mathrm{const} = 0.2$ for all three models. The $y$-axis is on a long scale in both panels.}
    \label{fig:gtau}
\end{figure}

Next, Fig.~\ref{fig:bondcorr_vs_T} examines how the $S_\mathrm{B}(q = \pi)$ bond correlations observed at half-filling evolve with temperature. At very high temperatures, the system is dominated by thermal fluctuations. The details of the microscopic \gls*{eph} interaction matter very little at this temperature, and the bond correlations' strength is comparable across all three models. As $T$ decreases, the strength of the bond correlations grows, as expected. The value of $\chi_\mathrm{B}(\pi)$ in the acoustic and optical models is identical within error bars as a function of $T$, while the correlations in the bond model are consistently weaker. We can conclude from this data that the equivalency of the \gls*{oSSH} and \gls*{aSSH} models persists across all simulated temperatures and that these models produce stronger bond-correlations compared to the \gls*{bSSH} model for a fixed $\lambda^\mathrm{const}$. 

Figure~\ref{fig:gtau} provides an additional comparison of the imaginary time dependence of the local Green's function $G(r = 0,\tau)$ (Fig.~\ref{fig:gtau}a) and bond correlations $S_\mathrm{B}(q = \pi,\tau)$ (Fig.~\ref{fig:gtau}b). Again, the acoustic and optical \gls*{SSH} models produce identical results within error bars for both quantities, while the bond model has quantitatively different results.

\subsection{Single-particle spectral functions}\label{sec:Akw}

\begin{figure*}[t]
    \centering
    \includegraphics[width=\textwidth]{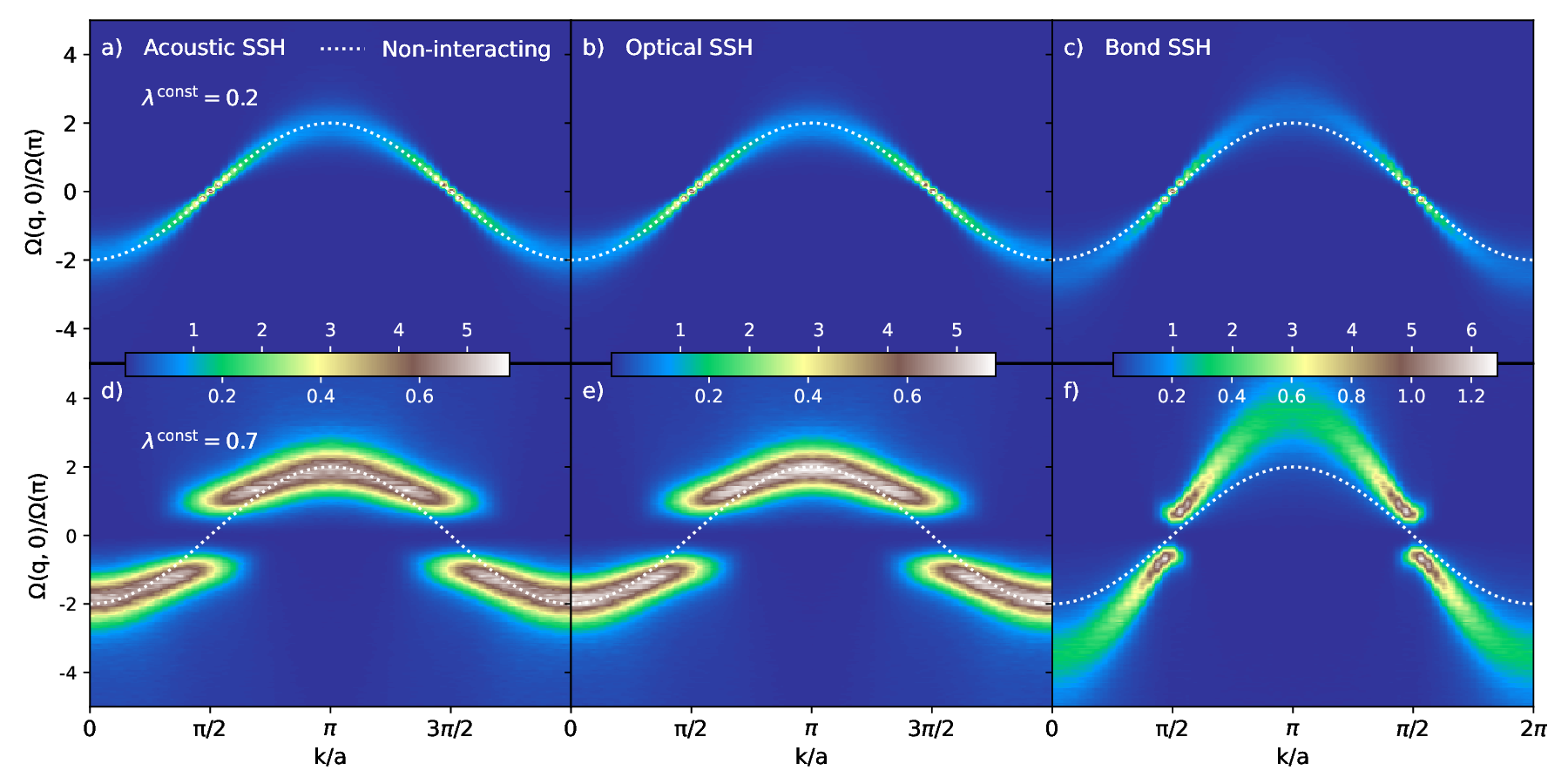}
    \caption{The single-particle spectral function $A(k,\omega)$ for the  
    various SSH models. The top row shows results for 
    the a) acoustic , b) optical , and c) bond SSH models with $\lambda^\mathrm{const} = 0.2$. The bottom row shows results for the same models but this time with $\lambda^\mathrm{const} = 0.7$. The results in each panel were obtained on $L=64$ site chains at half-filling $\langle n\rangle = 1$ ($\mu = 0$) with $\Omega_\mathrm{b}=\Omega_\mathrm{o}=2\Omega_\mathrm{a}=0.1$ and $\beta = 15/t$. The white dashed lines in each panel indicate the non-interacting dispersion at half-filling $\epsilon(k)= -2t\cos(ka)$.}
    \label{fig:Akw}
\end{figure*}

We now turn to the spectral properties of the three models in both the weak and strong coupling limits. Fig.~\ref{fig:Akw} plots the single particle spectra function $A(k,\omega)$ for the models for fixed $\lambda^\mathrm{const} = 0.2$ (top row) and $\lambda^\mathrm{const} = 0.7$ (bottom row). We have performed the calculations on long $L = 64$ site chains to achieve fine momentum resolution, and fixed the inverse temperature to $\beta = 15/t$.  

At weak coupling [Figs.~\ref{fig:Akw}a-c], all three models produce spectra that disperse through the Fermi level, indicative of a metallic phase at this temperature. All three spectra exhibit canonical signatures of the $\gls*{eph}$ interaction. Specifically, the peaks broaden as the quasiparticle dispersion crosses the phonon energy $\omega = \pm \Omega(q=\pi)$, which is a common characteristic of \gls*{eph} coupled systems. This broadening would usually be accompanied by a kink in the band dispersion at $\omega = \Omega(\pi)$~\cite{Engelsberg1963coupled, Nosarzewski2021spectral}, but we are unable to resolve such a feature in our data. This may be due to the low values of the coupling and phonon energy $[\Omega(\pi)/t = 0.1]$ or difficulties in resolving such a subtle spectral feature in the analytically continued data. Notably, the spectra for the \gls*{aSSH} and \gls*{oSSH} models are indistinguishable, consistent with the equivalence described in the previous sections, while the bSSH model [Fig.~\ref{fig:Akw}c] has increased bandwidth $W \approx 4.6t$. This latter observation is consistent with the effective bandwidth $W_\mathrm{eff} = 4t_\mathrm{eff} = 5t$ one would estimate using a mean-field-like analysis of the effective hopping $t_\mathrm{eff} = t-\alpha_\mathrm{b}\langle X\rangle \approx 1.253$ ($\alpha_\mathrm{b} =0.0447$ and $\langle X \rangle = -5.666$, see Fig.~\ref{fig:deltaL} and Tbl.~\ref{tbl:coupling}). 

Turning now to $\lambda^\mathrm{const} = 0.7$ [Figs.~\ref{fig:Akw}d-e], we find that all three spectra are significantly broadened and open a gap at the Fermi level, indicative of \gls*{BOW} insulating state. 
As with the weak coupling case, the spectra for the \gls*{aSSH} and \gls*{oSSH} models are identical and have a bandwidth comparable to the non-interacting value. Conversely, the \gls*{bSSH} model's spectra are sharper (note the difference in the intensity scale) and have a smaller gap, indicating that the \gls*{BOW} correlations and quasi-particle dressing of the carriers are weaker in the bond model compared to the other two. The observed bandwidth of the \gls*{bSSH} model increases to $W \approx 7.3t$ for this value of the coupling, which is again consistent with the estimate $W_\mathrm{eff} = 4t_\mathrm{eff} = 7.45t$ ($\langle X\rangle = -10.335$ and $\alpha_\mathrm{b} = 0.0836$) obtained from a mean-field-like analysis. 

\subsection{Renormalized Phonon Dispersions}\label{sec:Bqnu}
\begin{figure*}[t]
    \centering
    \includegraphics[width=\textwidth]{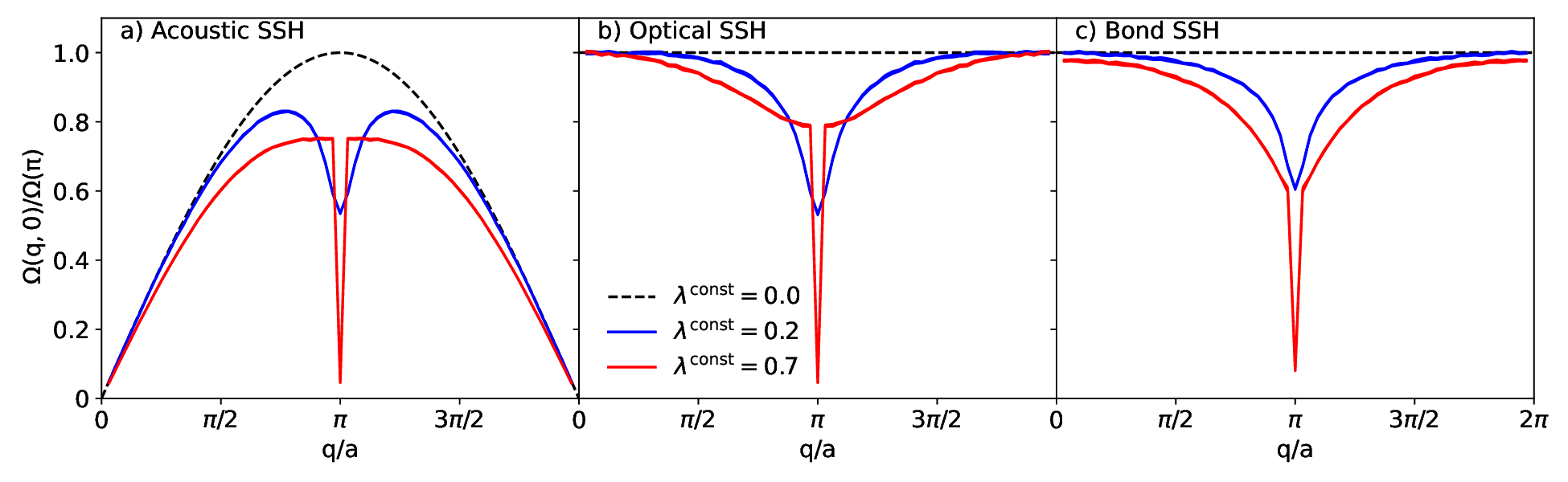}
    \caption{The renormalized phonon dispersion [see Eq.~\eqref{eq:phonon_disp}] for the 
    a) acoustic, b) optical, and c) bond SSH models with $\lambda^\mathrm{const} = 0$, $0.2$, and $0.7$. All results were obtained on $L=64$ site chains at half-filling $\langle n\rangle = 1$ ($\mu = 0$) with $\Omega_\mathrm{b}=\Omega_\mathrm{o}=2\Omega_\mathrm{a}=0.1$ and $\beta = 15/t$.
    }
    \label{fig:phonon_dispersion}
\end{figure*}

Figure~\ref{fig:phonon_dispersion} plots the renormalized phonon dispersion relations 
$\Omega(q,\mathrm{i}\nu_n = 0)$ [see Eq.~\eqref{eq:phonon_disp}] for the same parameters used in the previous section. At weak coupling ($\lambda^\mathrm{const} = 0.2$), the dispersion relations are weakly renormalized, with the degree of mode softening increasing as $q$ approaches the Brillouin zone boundary. The observed softening reflects the formation of BOW correlations driven by FS nesting and the underlying momentum dependence of the \gls*{eph} coupling, which is strongest near the zone boundary [see Eq.~\eqref{eq:fkq}]. However, the $q=\pi$ mode not softening to zero at this temperature is consistent with the absence of a gap in the corresponding single-particle electron spectral function. 

Similar $q$-dependent normalizations are observed for strong coupling ($\lambda^\mathrm{const} = 0.7$); however, the broad dip around the zone boundary becomes shallower in the optical model compared to the weak coupling result. At the same time, a sharp discontinuity appears at $q = \pi$, where this mode softens nearly to zero. For this value of the coupling, the system is in an insulating, dimerized state with large lattice displacements. 
We attribute the finite value of $\Omega(\pi,0)$ to a finite size effect~
\cite{CohenStead2023hybrid}. Interestingly, for both weak and strong coupling we find that $\Omega_\mathrm{a}(\pi,0) = \Omega_\mathrm{o}(\pi,0)\ne \Omega_\mathrm{b}(\pi,0)$. In addition, the renormalized dispersion for the \gls*{bSSH} model also does not approach its non-interacting value at $q \rightarrow 0$, which reflects the fact that this model has a non-zero coupling to this mode. These behaviors are fully consistent with the (in)equivalences between the respective models discussed in the previous sections. 

Finally, we remark that we recover the expected linear dispersion for the \gls*{aSSH} model as $q\rightarrow 0$, which demonstrates that our \gls*{HMC} sampling algorithm correctly captures the long-wavelength behavior of this phonon branch.  

\section{Discussion}
Our results demonstrate that at half-filling, the \gls*{1D} acoustic and optical \gls*{SSH} models can be made to produce identical results within error bars for a suitable choice of phonon energies and \gls*{eph} coupling constant. One has to set the phonon energy scale and microscopic \gls*{eph} coupling constants such that energy of the $q = 2k_\mathrm{F} = \pi$ modes and mode-resolved dimensionless coupling for scattering across the Fermi surface $|g(k_\mathrm{F},2k_\mathrm{F})|^2/\Omega(2k_\mathrm{F})$ are equal for both models. For this choice of parameters, both models exhibit the same $q = \pi$ lattice dimerization, where alternating bonds expand and contract along the length of the chain. 

The equivalence between the \gls*{aSSH} and \gls*{oSSH} models in \gls*{1D} may be expected by considering the various energies entering the respective Hamiltonians. The potential energy cost for this dimerization is identical for both models when $\Omega_\mathrm{o}(\pi) = \Omega_\mathrm{a}(\pi)$. The dimerization also couples to the electronic hopping integrals via the same microscopic \gls*{eph} interactions in both models. It is unsurprising that they produce similar physics in the weak coupling limit when viewed in this light. Conversely, we found that the \gls*{bSSH} model differs significantly from the \gls*{aSSH} and \gls*{oSSH} models, which we attribute to the non-zero coupling to the $q = 0$ phonon mode in the former case. This same coupling is also reflected in an overall contraction of the lattice observed in the \gls*{bSSH} model, which is forbidden in the \gls*{aSSH} and \gls*{oSSH} models. 

Our results do not agree with Ref.~\cite{Weber2015excitation}, which found an equivalence between the bond and acoustic models at half-filling. The origin of this discrepancy is unclear, but we suspect it may be related to how that work treated the coupling to the $q = 0$ modes in these models. 

The equivalency here has been established only for the half-filled \gls*{1D} model. Upon doping, we found that the acoustic and optical models agreed only qualitatively when we fixed the model parameters based on the Fermi surface of the half-filled model. We suspect it is possible to retain the consistency between the \gls*{aSSH} and \gls*{oSSH} models by adjusting the phonon dispersion and \gls*{eph} coupling value to reflect the Fermi momentum of the doped system. However, maintaining this equivalence in higher dimensions, where the Fermi surface may no longer be well nested, would be challenging. Additionally, we have found that in the \gls*{bSSH} mode the coupling to the ${\bf q} = 0$ mode in the bond model persists in our simulations of this model in 2D. Therefore, we expect this model to remain formally inequivalent to the optical and acoustic \gls*{SSH} models in higher dimensions. We conclude that the bond, optical, and acoustic variants of the \gls*{SSH} models are in general quite different models, and this aspect should be kept in mind when drawing broader conclusions from results obtained from one of these models. \\

\section*{acknowledgments}
This work was supported by the U.S. Department of Energy, Office of Science, Office of Basic Energy Sciences, under Award Number DE-SC0022311. This research used resources of the Oak Ridge Leadership Computing Facility, a DOE Office of Science User Facility supported under Contract No. DE-AC05-00OR22725.

\appendix
\section{Sign changes in the effective hopping integral}\label{sec:sign_change}

The modulation of the hopping integrals by the lattice produces an effective hopping that now depends on the phonon displacement $\hat{X}_{{\bf i},\nu}$ such that
\begin{equation}
    \begin{aligned}
    t^{\mathrm{a(o)}}_{{\bf i},\nu}&= t - \alpha_\mathrm{a(o)}\left\langle\hat{X}_{{\bf i}+{\bf a}_\nu,\nu}-\hat{X}_{{\bf i},\nu}  \right\rangle,~\mathrm{and}\\
    t^\mathrm{b}_{{\bf i},\nu}&= t - \alpha_\mathrm{b}\left\langle\hat{X}_{{\bf i},\nu}\right\rangle. 
    \end{aligned}
\end{equation}
Crucially, the sign of the effective hopping can change in all three cases if the lattice displacements are large enough. These unphysical sign changes lead to dimerization via a mechanism utterly distinct from the standard Fermi-surface-nesting scenario, as discussed in Ref.~\cite{banerjee2023ground}. 
\begin{figure}[b]
    \centering
    \includegraphics[width=\columnwidth]{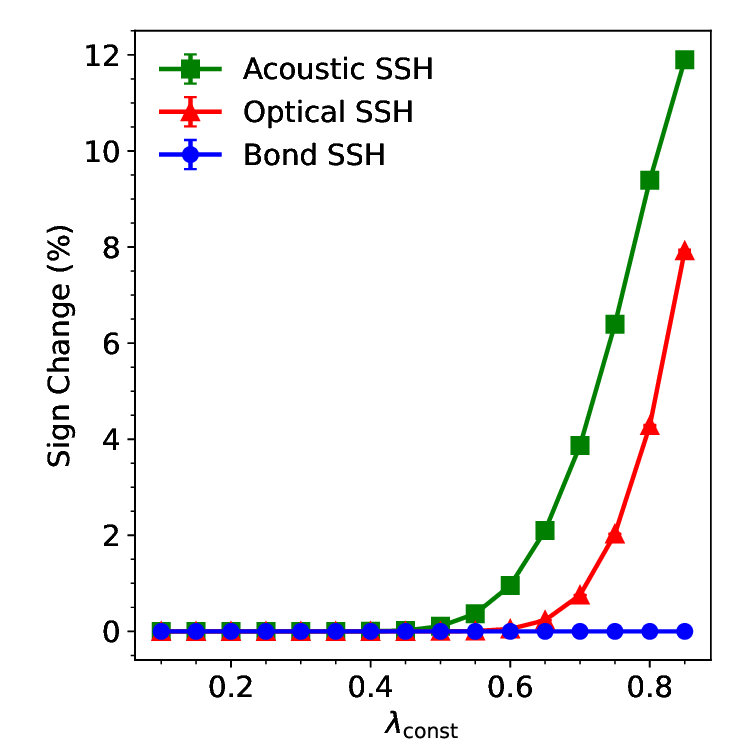}
    \caption{The percentage number of times that the effective hopping integral changes sign during our \gls*{DQMC} simulations as a function of $\lambda^{\mathrm{const}}$. Results were obtained on an $L = 24$ site chain at an inverse temperature $\beta=15/t$ and a fixed $\langle n \rangle = 1$ ($\mu = 0.0$).}
    \label{fig:sign_change}
\end{figure}
Our HMC sampling procedure can produce lattice configurations where the sign of the hopping has changed. Therefore, we monitored the percentage of times this occurs in our simulations. Figure~\ref{fig:sign_change} presents results for a half-filled $L = 24$ site chain as a function of $\lambda^\mathrm{const}$ and a fixed $\beta =15/t$. Note that we do \textit{not} average the phonon positions over imaginary time before calculating this percentage. The results indicate that the hopping integrals rarely change in our simulations for $\lambda^\mathrm{const} \le 0.5$. Some sign switching does occur in both the acoustic and optical models $\lambda^\mathrm{const} \gtrsim 0.6$, however, it is still fairly rare in the \gls*{DQMC} simulations. For example, the \gls*{aSSH} model has a sign switching of approximately 4\% at $\lambda^\mathrm{const} = 0.7$. Both the \gls*{aSSH} and \gls*{oSSH} models become more prone to these unphysical sign changes in the strong coupling limit, while the \gls*{bSSH} model is not. We attribute this difference to the coupling to the $q = 0$ mode in the latter case, which causes the entire lattice to contract, thus increasing $|t^\mathrm{b}_{{\bf i},\nu}|$.

\begin{figure}[t]
    \centering
    \includegraphics[width=\columnwidth]{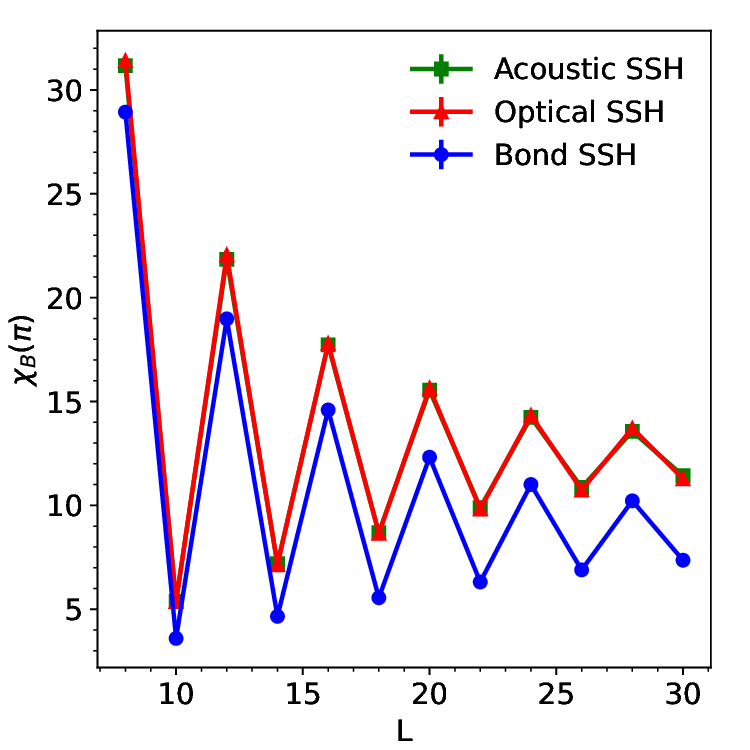}
    \caption{The bond susceptibility $\chi_\mathrm{B}(\mathrm q=\pi)$ of the $\langle n \rangle = 1$ ($\mu = 0.0$) models as a function of the chain length. Results were obtained at an inverse temperature of $\beta=15.0$ and a fixed dimensionless coupling $\lambda^{\rm{const}}=0.2$.}
    \label{fig:bondsusVSL}
\end{figure}

\section{Finite size effects at half-filling}\label{sec:finite_site}

Figure~\ref{fig:bondsusVSL} presents the dependence of the bond susceptibility $\chi_\mathrm{B}(\mathrm q=\pi)$ for a half-filled system as a function of chain length $L$. Results are shown for the \gls*{aSSH}, \gls*{bSSH}, and \gls*{oSSH} models for a fixed $\beta = 15/t$ and $\lambda^\mathrm{const} = 0.2$. In this case, we observe a clear oscillation in the value of $\chi_\mathrm{B}(\mathrm q=\pi)$, and the strength of the correlations are under- (over-) predicted relative to the thermodynamic limit whenever the chain length is $L = 4n$ ($2n$), where $n$ is an integer.  Importantly, all three models exhibit the same oscillatory behavior with $L$, and both the \gls*{aSSH} and \gls*{oSSH} models produce identical values of $\chi_\mathrm{B}(\pi)$ for a given chain length. Thus, while our $L=24$ chains are short enough to retain some cluster size dependence, our conclusions on the equivalence of the three models will hold in the thermodynamic limit. 

\bibliography{references}

\end{document}